\documentclass[preprint,12pt]{elsarticle}

\usepackage{graphicx}

\usepackage{amssymb}

\journal{Nuclear Instruments and Methods in Physics Research A}

\begin{document}

\begin{frontmatter}

  \title{A Feasibility Study for the Detection of Supernova Explosions with an Undersea Neutrino Telescope.}

 \author{A.~Leisos\corref{cor1}}
  \cortext[cor1]{Corresponding author.~Tel: +302610367523; Fax: +302610367528}
  \ead{leisos@eap.gr}
  \author{A.G.~Tsirigotis}
  \author{and S.E.~Tzamarias}
  \author{on behalf of the KM3NeT consortium}
  \address{Physics Laboratory, School of Science \& Technology, Hellenic Open University, Tsamadou 13-15 \& Ag.~Andreou, Patras 26222, Greece}

\begin{abstract}
We study the potential of a very large volume underwater Mediterranean neutrino telescope to observe neutrinos from supernova (SN) explosions within our galaxy. The intense neutrino burst emitted in a SN explosion results in a large number of MeV neutrinos inside the instrumented volume of the neutrino telescope that can be detected (mainly) via the reaction $\bar{\nu}_{e} + p \longrightarrow e^{+} + n$. In this study we simulated the response of the underwater neutrino telescope to the electron antineutrino flux predicted by the Garching model for SN explosions. We assumed that the neutrino telescope comprises 6160 direction sensitive optical modules, each containing 31 small photomultiplier tubes. Multiple coincidences between the photomultiplier tubes of the same optical module are utilized to suppress the noise produced by $^{40}K$ radioactive decays and to establish a statistical significant signature of the SN explosion.
\end{abstract}

\begin{keyword}
  Supernova \sep Neutrino Telescope 
  \PACS 95.55.Vj
\end{keyword}

\end{frontmatter}
\section{Introduction}
Core collapse supernovae produce huge bursts of MeV neutrinos which carry most of the gravitational energy ($\sim 99\%$) of a massive star in a timescale of a few seconds. 
The released energy ($\sim 10^{53}$~ergs) is expected to be roughly equipartioned among the different flavors of neutrinos\footnote{About 1\% of this energy is emmited in the form of electron neutrinos from an initial neutrinization burst lasting a few ms.} emerging promptly from the dense core, a few hours before the optical signal from the stellar envelope. 
The neutrino burst's time, flavor and energy structure carry information about the explosion mechanism \cite{Raffelt} and they can also be used to study  vacuum and matter neutrino oscillations \cite{Dighe}, the neutrino mass hierarchy \cite{Smirnov} or even exotic physics phenomena related to large extra dimensions and couplings to axions \cite{Hann}. 
Since the SN1987A explosion when 24 neutrinos were detected by the Kamiokande-II, IMB and Baksan, many other experiments were designed \cite{Scholberg, nostos} to observe and study SN neutrinos. 
These detectors can also provide an early alarm for a SN explosion offering the chance to the astronomical community to make unprecedented observations of the early turn-on of the supernova light curve. 
For this purpose a SuperNova Early Warning System (SNEWS \cite{snews}) already exists providing also  a trigger to detector facilities  which are not able to trigger by themselves on a supernova signal.

At the time of this writing, supernova neutrinos are the only neutrinos detected beyond our solar system. However, a lot of effort has been invested in observing high energy neutrinos from galactic and extragalactic sources.
Very large volume neutrino telescopes are designed \cite{km3net} and are already operating \cite{icecube}. Such telescopes are consisting of thousands Photomultiplier Tubes (PMT) and instrumented volumes of the order of  km$^{3}$.
Neutrino interaction products, mainly muons and electrons, are detected through the Cherenkov light produced  in water or in ice, providing  a neutrino detection energy-threshold of about 100~GeV. 
However, it has been shown \cite{icepapers1,icepapers2,icepapers4,icepapers5,novel} that bursts of low energy (MeV) SN neutrinos are detectable  in neutrino telescopes in ice, by observing a global rise of the PMTs counting rates which exceeds significantly the background  rate.  
 Since undersea  neutrino detectors are suffering from background noise, due to the $^{40}K$ natural radioactivity of the sea-water, it was generally believed that only very dense detector configurations could observe SN signatures \cite{dumand}.
In this paper we demonstrate that deep undersea neutrino telescopes equipped with optical sensors that comprise many small PMTs, are capable of detecting burst of SN neutrinos without imposing any restrictions on the detector architecture.
\section{The Underwater Mediterranean $\nu$-Telescope}
In this study, we assume an underwater neutrino telescope following the Mediterranean  KM3NeT architecture \cite{km3net}. 
The KM3NeT detector will consist of several hundreds of vertical structures (Detection Units - DUs), which carry photo-sensors and devices for calibration and environmental measurements, arranged vertically on Storeys. 
Each Storey will support two photo-sensors. The photo-sensor unit is a digital optical module (DOM) consisting of a 17 inch diameter pressure resistant glass sphere, housing 31 3-inch small photomultiplier (small-PMT) tubes, their high-voltage bases and their interfaces to the data acquisition system with nanosecond timing precision \cite{km3netppm}.
The segmentation of the photocathode area in such a Multi-PMT DOM aids in distinguishing single-photon from multi-photon hits, providing thus a better background rejection and trigger efficiency.
The front-end electronics will utilize time over threshold (TOT) measurements that carry amplitude and timing information, which allow for reconstruction of the original signal.
The readout system will support the broadcasting of all the digitized data to the shore.  

In this work, we studied the SN discovery potential of such a $\nu$-Telescope configuration, deployed in a Mediterranean site of 3500~m depth, consisting of 154 DUs, with a mean horizontal separation of 180~m. Each DU was arranged in 20 storeys (6~m long bars), with a vertical distance between them of 40~m.  Each storey carries  two photo-sensors, one at each end of the bar. 
\section{Discovery Potential for SN explosions}
Assuming that the expected number of background events in a time interval $T$ is  $\mu_{\mathrm{bck}}(T)$ and that $m(T)$ events have been observed in this time period, the condition that there is less than $5.72 \times 10^{-7}$ 
probability the observed events  to be produced solely by the fluctuating  background is:
\begin{equation}
1-\sum_{i=0}^m \frac{\mu_{\mathrm{bck}}^i(T) e^{-\mu_{\mathrm{bck}}(T)}}{i!}<1-erf(\frac{5}{\sqrt{2}})
\label{eq_m}
\end{equation}
In case that $m(T)$ is large, the above formula can be replaced by its Gaussian approximation, $s(T)>5\sqrt{\mu_{\mathrm{bck}}}$, where $s(T)=m(T)-\mu_{\mathrm{bck}}(T)$ is the excess of events with respect to the expected background. Obviously, the aim of this study is to define an analysis strategy in order to maximize the following quantity:
\begin{equation}
r=\frac{s(T)}{\sqrt{\mu_{\mathrm{bck}}(T)}}
\label{eq_DP}
\end{equation}
In the case that the excess  $s(T)$ is due to neutrinos from a SN explosion, it will depend on the proximity of the SN to earth, i.e.~it will be inverse proportional to the squared distance  ($d^{2}$).
Consequently, the  $5\sigma$ Confidence Level (CL) requirement in (\ref{eq_m}) determines the maximum distance of a typical SN in order to be detected, or equivalently, given the distance of the SN, the statistical significance of the observation  can be evaluated.
\section{SN model and neutrino interactions}
The SN explosion model used in this work, is the so called ``Garching Model'' \cite{garching} which assumes an initial star with mass of 8-10 solar masses, which represent 30\% of all the studied SNe.

Among the interactions of these SN neutrinos in water, inside the instrumented volume, we consider only the inverse beta decay (IBD) of the electron antineutrino  ($\bar{\nu}_{e} + p \longrightarrow e^{+} + n$) which is the most dominant process\footnote{The neutrino energy threshold for the IBD process is 1.8~MeV, accounting for about 95\% of the signal.}. We used the IBD total cross section $\sigma (E_{\bar{\nu}_{e}})$ as well as the differential cross sections $$ \frac {d \sigma} {dE_{e^{+}}} (E_{ \bar{\nu}_{e} },E_{e^{+}}),
\frac {d \sigma} {d\mathrm{cos}\theta} (E_{ \bar{\nu}_{e} },\mathrm{cos}\theta)$$ (where $\theta$ is the angular difference between the positron direction and the neutrino direction) from  \cite{strumia}, which are very accurate for the energy range of SN neutrinos\footnote{NNLO corrections are less than 1\%.}. 

The time-integrated electron antineutrino differential flux  (integrated for an observation time interval, $T$, which starts with the bounce) is expressed as: 
$$\frac{d\Phi_{\bar{\nu}_{e}}}{dE_{\bar{\nu}_{e}}}=\frac{1}{4\pi d^{2}} \int_0^{T}\frac{L_{\bar{\nu}_{e}}(t)}{\langle E_{\bar{\nu}_{e}} \rangle (t)} f(E_{\bar{\nu}_{e}};\langle E_{\bar{\nu}_{e}} \rangle,a) dt$$
where $L_{\bar{\nu}_{e}}$ and $E_{\bar{\nu}_{e}}$ are the luminosity and the energy of the antineutrino respectively, whilst $d$ is the distance of the SN from the earth. 
$f(E_{\bar{\nu}_{e}};\langle E_{\bar{\nu}_{e}} \rangle,a)$ is the energy spectrum of the electron antineutrino as parametrized in \cite{keil}, where $\langle E_{\bar{\nu}_{e}} \rangle$ is the mean energy of the neutrino and $\alpha$ a parameter representing the amount of spectral pinching. 
The time evolution of $L_{\bar{\nu}_{e}}$, $\langle E_{\bar{\nu}_{e}} \rangle$ and  $\alpha$ are given by the  Garching Model.
The integration time $T$ is chosen as in \cite{novel} in order to maximize the discovery potential for SN detection (i.e.~Eq.~(\ref{eq_DP})) and it is found to be $T=2.95~\mathrm{s}$, using the time structure of the neutrino burst predicted by the Garching Model.
\section{Simulation studies}
The positron emitted in SN neutrino induced IBD process is sufficiently energetic (mean energy 17.19~MeV) to produce Cherenkov radiation in the sea water, which is detectable by the photo-sensors of a neutrino telescope. The response of the detector to the IBD positrons and background processes is simulated by utilizing the HOURS \cite{hours} package (Hellenic Open University Reconstruction and Simulation package), which describes in detail all the relevant physical processes (e.g.~secondary interactions, propagation of Cherenkov photons), the response of the PMTs and the digitization electronics. Using the simulated events we evaluated the effective volume, $V_{\mathrm{eff}}$, for IBD interactions  and the interaction density, $\rho_{\mathrm{int}}(d;T)$, which expresses the number of IBD neutrino interactions per unit volume, occurring inside an observation time interval $T$,  caused by the explosion of a SN at distance $d$. The effective volume is estimated as the water volume, around a DOM, in which we have simulated IBD positron production, weighted by the fraction of the interactions observed by the DOM, under certain observation criteria. The interaction density is found by estimating the integral,
$$\rho_{\mathrm{int}}(d;T)=\rho_{\mathrm{target}}\int_{0}^{\infty}\frac{d\Phi_{\bar{\nu}_{e}}}{dE_{\bar{\nu}_{e}}}\sigma(E_{\bar{\nu}_{e}})dE_{\bar{\nu}_{e}}$$ by Monte Carlo integration, where $\rho_{\mathrm{target}}$ is the free proton density in sea water at the detector deployment depth and the cross sections are taken as described in Section 4.
As an example, for a SN explosion at a distance of $d=10~\mathrm{kpc}$ and for $T=2.95~\mathrm{s}$ the interaction density at a depth of 3500~m is estimated to be $\rho_{\mathrm{int}}=0.078~\mathrm{m}^{-3}$. Finally  the expected number of  observed  IBD interactions, from a SN explosion at a distance $d$ is evaluated as,
\begin{equation}
 N_{\mathrm{obs}}=\rho_{\mathrm{int}}(d;T) \times V_{\mathrm{eff}}
\label{obs}
\end{equation}
\begin{figure}
\centering
\includegraphics*[width=10.0cm]{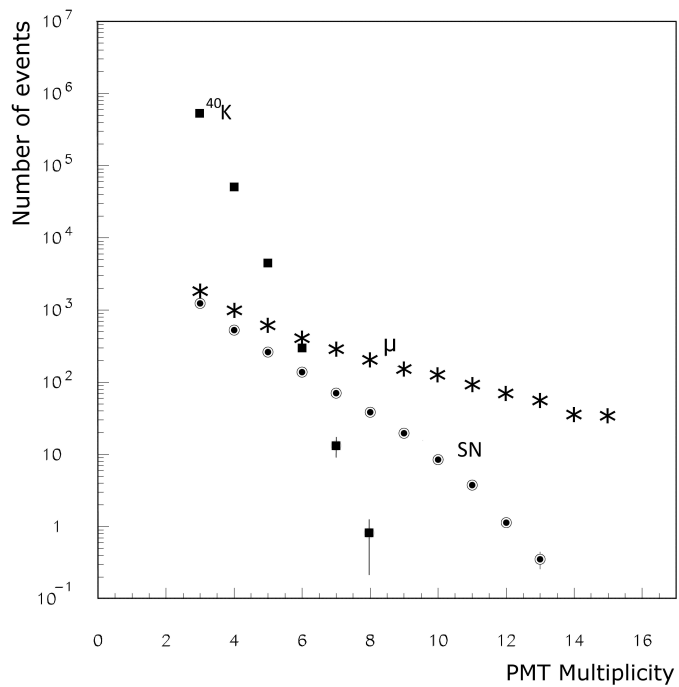}
\caption{The expected number of observed events (active DOMs), in an observation time interval $T=2.95~\mathrm{s}$, as a function of the PMT multiplicity for three type of events: $^{40}K$ decays, atmospheric muons and SN neutrinos. The SN explosion is assumed to occur at a distance of 10~kpc and the $\nu$-Telescope is assumed to  be deployed at a depth of 3500~m containing 6160 DOMs.}
\label{fig_events}
\end{figure} 

As already mentioned, the main background source to SN neutrinos is the natural radioactivity of $^{40}K$ in the sea water. 
The $^{40}K$ decay rate is about $13.6 \times 10^{3}~\mathrm{m^{-3}~s^{-1}}$, five orders of magnitude greater than the expected IBD rate from a SN at 10~kpc. 
However,  the  released $^{40}K$ decay energy   is relatively small\footnote{the mean electron energy is 1.27~MeV and the energy of the photon due to electron capture is about 1.46~MeV} with respect to the energy of the SN neutrino induced IBD positron. 
The effective volume corresponding to the $^{40}K$ decays, is thus much smaller than the effective volume for SN neutrino IBD interactions. Furthermore the products of $^{40}K$ decays have less probability to activate synchronously several small PMTs of the same DOM. We define thus, as the main observation criterion, the PMT multiplicity, which is  the number of PMTs in the same DOM that are simultaneously active within a coincidence time window of 100~ns. 
 Fig.~\ref{fig_events} presents the total number of active DOMs due to $^{40}K$ decays, collected during an observation time interval of $T=2.95~\mathrm{s}$,  as a function of the PMT multiplicity, in comparison with the  corresponding number due to SN neutrino induced IBDs, assuming that the SN explosion occurred at a distance of 10~kpc.
Although the background due to  $^{40}K$ decays is significantly reduced  at high  PMT multiplicities, it is expected that 
 atmospheric muons will still  contribute to the background up to very high PMT multiplicities, due to the causal connection between hits produced by muons. In order to quantify this muonic background we have used HOURS with generation input from the MuPage package \cite{mupage} with  an energy threshold as low as 50~GeV. 
The contribution of this type of background  is also shown as a function of the PMT multiplicity   in Fig.~\ref{fig_events}. 
\begin{figure}
\centering
\includegraphics*[width=10.0cm]{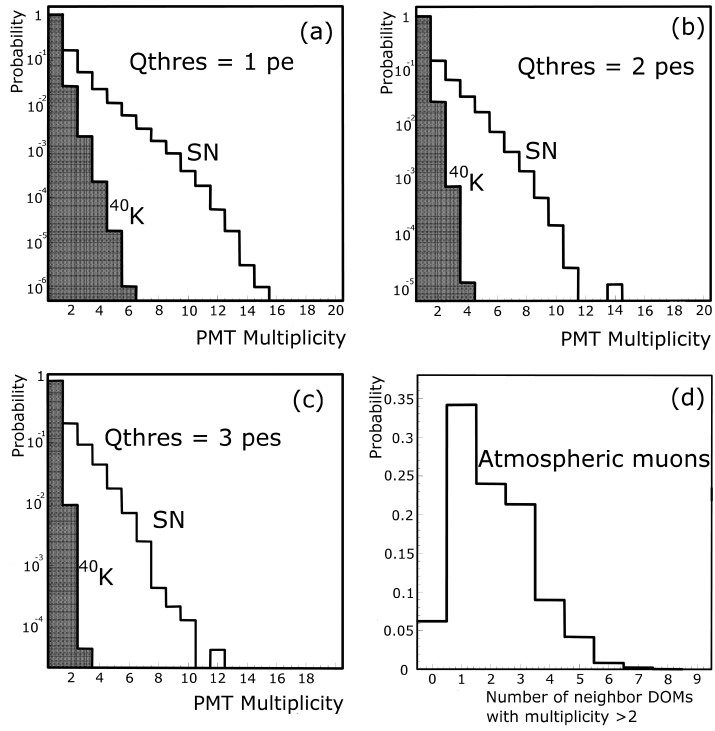}
\caption{The probability of a DOM to exhibit a certain PMT multiplicity for three different small-PMT pulse-amplitude thresholds: $Q_{\mathrm{thres}}=1~\mathrm{pe}$ (a), $Q_{\mathrm{thres}}=2~\mathrm{pes}$ (b) and $Q_{\mathrm{thres}}=3~\mathrm{pes}$ (c). (d): The probability that a DOM activated by an atmospheric muon, with PMT multiplicities greater that 5, has simultaneously (within a time window of 100~ns) $N$ active neighboring DOMs, with multiplicity greater than 2.}
\label{fig_cuts}
\end{figure} 
\section{Observation Strategy and Results}
As shown in Fig.~\ref{fig_events}, a high  PMT multiplicity rejects the  $^{40}K$ background.
A further suppression of this background can be achieved when  we apply selection criteria, based on the amplitude of the small-PMT signals.
Fig.~\ref{fig_cuts} (a,b and c) shows the probability of a DOM, activated by a $^{40}K$ decay or SN neutrino induced IBD, to exhibit a certain PMT multiplicity, for three different small-PMT pulse-amplitude thresholds, $Q_{\mathrm{thres}}$. As an example, in Fig.~\ref{fig_cuts}(a) a small PMT is considered active when its pulse amplitude exceeds an amplitude-threshold corresponding to the average pulse amplitude produced by one photoelectron (pe), $Q_{\mathrm{thres}}=1~\mathrm{pe}$. The probability of the  $^{40}K$ background to produce high PMT multiplicity events decreases faster than the signal, as the threshold requirement is getting higher. However we found that this requirement has not any significant effect on the background produced by atmospheric muons. 
On the other hand, it is very improbable that the  $^{40}K$ background and the SN neutrino induced IBDs  will  activate small PMTs in different DOMs (especially when high PMT multiplicity is required). On the contrary, muon tracks produce very often time-correlated hits in different DOMs, with high PMT multiplicity.
This is demonstrated  in Fig.~\ref{fig_cuts}(d), where it is shown the probability that a DOM activated by an atmospheric muon, with PMT multiplicity greater that 5, has simultaneously (within a time window of 100~ns) $N$ active neighbor DOMs, with PMT multiplicity greater than 2. Rejecting thus all the active DOMs with PMT multiplicities greater than 5, when neighboring with at least one, simultaneously-active DOM, with PMT multiplicity at least 3, the atmospheric muon background is reduced by  94\% whilst only 10\% of the signal is lost.
\begin{figure}
\centering
\includegraphics*[width=10.0cm]{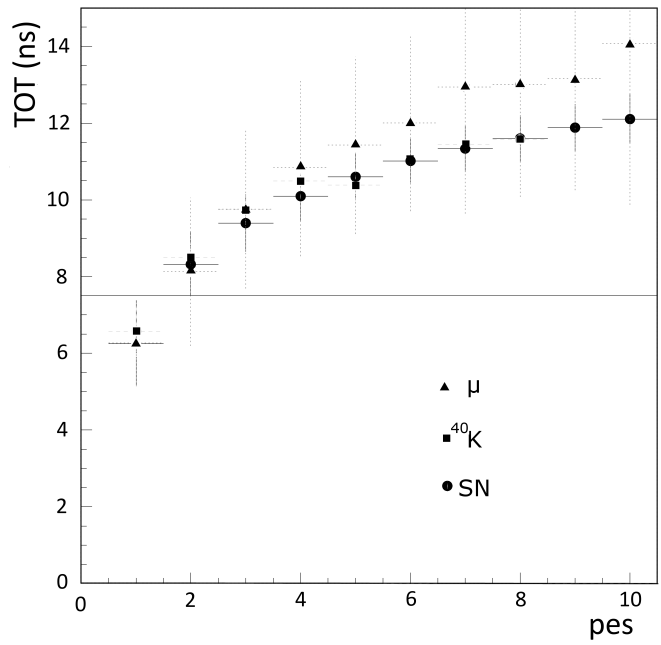}
\caption{The mean value of the TOT value with respect to the number of the photoelectrons for three type of events: IBD interactions, $^{40}K$ decays and atmospheric muon events. The error bars correspond to the RMS of the TOT distribution.}
\label{fig_pulse}
\end{figure}
We used fully simulated signal ($10^{7}$ SN neutrino induced IBDs) and background ($10^{11}$ $^{40}K$ and $10^{6}$ atmospheric muon) events in order to estimate (using Eq. (\ref{obs})) the number of signal events from a SN explosion at 10~kpc, as well as the expected number of $^{40}K$ and atmospheric muon background events, for various sets of event selection criteria. In the simulation we have assumed that the response function of the PMTs produced by a single pe is described by the function $$f(t)=a t^{b} e^{-\frac{t}{c}}$$ that the charge resolution is 35\%, the time jitter is 1~ns and the rise time of the pulse is 2.5~ns. We have also assumed that the digitization electronics provide the arrival time and the TOT of the pulse at the 30\% of the mean pulse amplitude corresponding to one pe. In Fig.~\ref{fig_pulse} it is shown  the mean value of the PMT-waveform  Time over Threshold  duration as a function of the corresponding  number of photoelectrons, for three type of events (IBD interactions, $^{40}K$ decays and atmospheric muon events). As an example, the condition  $TOT>7.5~\mathrm{ns}$  selects pulses corresponding to more than 1~pe.

The set of the optimal selection criteria, which are maximizing  the sensitivity of the detector as expressed in Eq.~(\ref{eq_DP}) are tabulated  in Table \ref{table}.
With these event selection criteria we expect 118.7 signal events from a SN explosion at 10~kpc on top of 0.8 $^{40}K$  and 15.4 atmospheric muon background events. For this expected background, an  observation with 5$\sigma$ significance requires, according to Eq.~(\ref{eq_m}), at least 39 total observed events (22.8 signal events). Consequently, the maximum distance of a SN explosion that could be observed by this detector is  estimated, by scaling proportionally to the squared distance, to be ($10~\mathrm{kpc} \times \sqrt{118.7/22.8}$) 22.8~kpc.
\begin{table}
\centering
\caption{Selection criteria that maximize the sensitivity of the detector.}
\begin{tabular}{|c|c|}
\hline
Description &  Optimum Value\\ 
\hline
\hline
PMT multiplicity $>$&  5 \\
\hline
$TOT_{\mathrm{threshold}}>$ & 7.5 ns\\ 
\hline
Number of PMTs (with $TOT>TOT_{\mathrm{thres}})>$&2\\ 
\hline
Neighbor DOMs (with PMT multiplicity $>$2) $<$ & 1\\
\hline
\end{tabular} 
\label{table}
\end{table}
\section*{Acknowledgment}
The KM3NeT project is supported by the EU in FP6 under Contract 140 no.~011937 and in FP7 under Grant no.~212525.
\bibliographystyle{elsarticle-num}

\end{document}